\documentclass[onecolumn,secnumarabic,amsmath,amssymb,balancelastpage,nofootinbib]{article}

\usepackage[e]{esvect}
\usepackage{color}         
\usepackage{graphics}      
\usepackage{graphicx}      
\usepackage{epsf}          
\usepackage{bm}            

\usepackage{natbib}
\usepackage{amssymb}
\usepackage{amsmath, esint}
\usepackage{mathrsfs}
\usepackage{framed}
\usepackage{bigints} 
\usepackage{enumitem}
\usepackage{pifont}
\usepackage[capbesideposition={left,center},facing=yes,capbesidewidth=8cm,capbesidesep=quad]{floatrow} 
\usepackage{setspace} 

\usepackage[none]{hyphenat} 

\usepackage[colorlinks=true]{hyperref}  

\definecolor{darkred}{rgb}{0.6,0,0}
\definecolor{darkgreen}{rgb}{0,0.5,0}
\definecolor{darkblue}{rgb}{0,0,0.6}
\hypersetup{ colorlinks,
linkcolor=darkblue,
filecolor=darkgreen,
urlcolor=darkgreen,
citecolor=darkred }

\newcommand{\noop}[1]{}

\begin{document}

\sloppy 

\bibliographystyle{authordate1}


\title{\vspace*{-35 pt}\Huge{How Electrons Spin}}
\author{Charles T. Sebens\\California Institute of Technology}
\date{July 26, 2024\\arXiv v.5\\\vspace{8 pt}Published version:\\\emph{Studies in History and Philosophy of Modern Physics} (2019), \textbf{68}, 40--50\\\vspace{8 pt} {\normalsize{[This is a post-publication version of the article with a\\note on further developments added at the end.]}}}

\maketitle
\begin{abstract}

There are a number of reasons to think that the electron cannot truly be spinning.  Given how small the electron is generally taken to be, it would have to rotate superluminally to have the right angular momentum and magnetic moment.  Also, the electron's gyromagnetic ratio is twice the value one would expect for an ordinary classical rotating charged body.  These obstacles can be overcome by examining the flow of mass and charge in the Dirac field (interpreted as giving the classical state of the electron).  Superluminal velocities are avoided because the electron's mass and charge are spread over sufficiently large distances that neither the velocity of mass flow nor the velocity of charge flow need to exceed the speed of light.  The electron's gyromagnetic ratio is twice the expected value because its charge rotates twice as fast as its mass.

\end{abstract}
\vspace*{12 pt}
\tableofcontents
\newpage

\section{Introduction}

In quantum theories, we speak of electrons as having a property called ``spin.''  The reason we use this term is that electrons possess an angular momentum and a magnetic moment, just as one would expect for a rotating charged body.  However, textbooks frequently warn students against thinking of the electron as actually rotating, or even being in some quantum superposition of different rotating motions.  There are three serious obstacles to regarding the electron as a spinning object:
\begin{enumerate}
\item Given certain upper limits on the size of an electron, the electron's mass would have to rotate faster than the speed of light in order for the electron to have the correct angular momentum.
\item Similarly, the electron's charge would have to rotate faster than the speed of light in order to generate the correct magnetic moment.
\item A simple classical calculation of the electron's gyromagnetic ratio yields the wrong answer---off by a factor of (approximately) 2.
\end{enumerate}
These obstacles can be overcome by taking the electron's classical state (the state which enters superpositions) to be a state of the Dirac field.  The Dirac field possesses mass and charge.  One can define velocities describing the flow of mass and the flow of charge.  The first two obstacles are addressed by the fact that the electron's mass and charge are spread over sufficiently large distances that the correct angular momentum and magnetic moment can be understood as resulting from rotation without either the velocity of mass flow or the velocity of charge flow exceeding the speed of light.  The electron's gyromagnetic ratio is twice the expected value because its charge rotates twice as fast as its mass.

In the next section I explain the three obstacles above in more detail.  Then, I consider how the obstacles are modified by the fact that some of the electron's mass is in the electromagnetic field that surrounds it.  The mass in the electromagnetic field rotates around the electron and thus contributes to its angular momentum.  Because the amount of mass in the electromagnetic field ultimately turns out to be small, this is not the dominant contribution to the electron's angular momentum.  But, the idea of mass rotating in a classical field appears again when we consider the Dirac field which describes the electron itself.  After an initial examination of this flow of mass and charge in the Dirac field, I show that the three obstacles can be overcome (in the manner described above) if we restrict ourselves to positive frequency modes of the Dirac field.  This restriction is imposed because negative frequency modes are associated with positrons in quantum field theory.  After presenting this account of spin, I compare it to other proposals as to how one might understand the electron's angular momentum and magnetic moment as arising from the motion of its mass and charge.

Before jumping into all of that, let me explain the focus on classical field theory in a paper about electron spin (a supposedly quantum phenomenon).  When one moves from classical field theory to a quantum description of the electron within the quantum field theory of quantum electrodynamics, the classical Dirac and electromagnetic fields are quantized.  Instead of representing the electron by a definite Dirac field interacting with a definite electromagnetic field, we represent the electron by a superposition of different field states---a wave functional that assigns amplitudes to different possible classical states of the fields.  The dynamics of this quantum state are determined by a wave functional version of the Schr\"{o}dinger equation and can be calculated using path integrals which sum contributions from different possible evolutions of the fields (different possible paths through the space of field configurations).  Seeing that the three obstacles above can be surmounted for each classical state makes the nature of spin in a quantum theory of those fields much less mysterious.  The electron simply enters superpositions of different states of rotation.

In the previous paragraph I assumed a ``field approach'' to quantum field theory where one starts from a relativistic classical theory of the Dirac field (with the Dirac equation giving the dynamics) and then quantizes this classical field theory to get a quantum theory of the Dirac field.  \citet{carrollblog} advocates such an approach.  He writes:
\begin{quote}
``What about the Klein-Gordon and Dirac equations? These were, indeed, originally developed as `relativistic versions of the non-relativistic Schr\"{o}dinger equation,' but that's not what they ended up being useful for. ... The Klein-Gordon and Dirac equations are actually not quantum at all---they are \emph{classical field equations}, just like Maxwell's equations are for electromagnetism and Einstein's equation is for the metric tensor of gravity. They aren't usually taught that way, in part because (unlike E\&M and gravity) there aren't any macroscopic classical fields in Nature that obey those equations.''\footnote{Carroll's point at the end of the quote is important (see also \citealp[chapter 8]{duncan}).  The motivation for studying the classical Dirac field in this paper is not that classical Dirac field theory emerges as an approximate description at the macroscopic level of what is happening microscopically according to quantum field theory.  The motivation is that quantum field theory describes superpositions of states of the classical Dirac field.  Classical Dirac field theory plays a key role in the foundations of quantum electrodynamics.}
\end{quote}
Carroll then goes on to explain that in quantum field theory the quantum state can be represented as a wave functional obeying a version of the Schr\"{o}dinger equation.

There is an alternative ``particle approach'' to quantum field theory where one begins instead from a relativistic quantum single particle theory in which the Dirac field is interpreted as a wave function for the electron and the Dirac equation gives the dynamics for that wave function.  In this theory, the electron is treated as a point particle in a superposition of different locations.  From this quantum particle theory, one can move to quantum field theory by transitioning from a relativistic single particle quantum theory to a theory with a variable number of particles.  Instead of having a wave function that assigns amplitudes to possible spatial locations for a single particle, one uses a wave function that assigns amplitudes to possible spatial arrangements of any number of particles (to points in the disjoint union of all $N$-particle configuration spaces).

These two approaches are often seen as different ways of formulating the very same physical theory.\footnote{For more on the field approach, see \citet{jackiw1990}; \citet[chapters 10 and 11]{hatfield}; \citet[chapter 4]{valentini1992}; \citet[section 12.4]{holland}; \citet{valentini1996}; \citet[chapter 3]{peskinschroeder}; \citet[chapter 4]{ryder}; \citet[pg.\ 241--242]{weinberg1999}; \citet{huggett2000}; \citet{wallace2001, wallace2006, wallace2017}; \citet[chapters 2 and 4]{tong}; \citet{baker2009}; \citet{struyve2010}; \citet{duncan}; \citet[section 4.3.1]{myrvold2015}.  For more on the particle approach, see \citet[chapters 6--8]{schweberQFT}; \citet[sec.\ 13.2]{bjorkendrellfields}; \citet{thaller1992}; \citet[chapter 3]{teller}; \citet[section 3]{durr2005}; \citet{deckert}, \citet{wallace2017}.  It is also possible to adopt a mixed approach.  For example, \citet{bohmhiley} take a field approach for bosons and a particle approach for fermions.}  However, in trying to understand what really exists in nature it is tempting to ask which approach better reflects reality.  Put another way:  Is quantum field theory fundamentally a theory of fields or particles?  This is a tough question and I will not attempt to settle it here (or even to develop the alternatives in much detail).  I seek only to display a single virtue of the first perspective: it allows us to understand electrons as truly spinning.  Adopting the first perspective is compatible with a number of different strategies for interpreting quantum field theory as it leaves open many foundational questions, such as:  Does the wave functional ever collapse?  Is there any additional ontology beyond the wave functional?  Are there many worlds or is there just one?

What follows is a project of interpretation, not modification.  It is generally agreed that the equations of our best physical theories describe an electron that has ``spin'' but does not actually rotate.  Here I present an alternative interpretation of the very same equations.  There is no need to modify these equations so that they describe a rotating electron.  Interpreted properly, they already do.

\section{The Obstacles}\label{obstacles}

The first obstacle to regarding the electron as truly spinning is that it must rotate superluminally in order to have the correct angular momentum.  One estimate for the radius of the electron is the classical electron radius, $\frac{e^2}{m c^2}\approx10^{-13}\mbox{cm}$ (which will be explained in the next section).  If you assume that the angular momentum of the electron is due entirely to the spinning of a sphere with this radius and the mass of the electron, points on the edge of the sphere would have to be moving superluminally \citep[problem 4.25]{griffithsQM}.  To get an angular momentum of $\frac{\hbar}{2}$ with subluminal rotation speeds, the electron's radius must be greater than (roughly) the Compton radius of the electron, $\frac{\hbar}{m c}\approx10^{-11}\mbox{ cm}$.  The relation between velocity at the equator $v$ and angular momentum $|\vec{L}|$ for a spherical shell of mass $m$ and radius $R$ is
\begin{equation}
|\vec{L}| = \frac{2}{3}m v R\ .
\end{equation}
Setting $|\vec{L}| = \frac{\hbar}{2}$ and $v=c$ then solving for $R$ yields a radius on the order of the Compton radius,
\begin{equation}
R=\frac{3}{4}\frac{\hbar}{m c}\ .
\end{equation}

Rejecting this picture of a spinning extended electron, one might imagine the mass of the electron to be confined to a single point.\footnote{The fact that there is mass in the electromagnetic field makes this difficult to imagine (see footnote \ref{whereisthemass}).}  If this were so, the electron's angular momentum---as calculated from the usual definition of angular momentum in terms of the linear momentum and displacement from the body's center of a body's parts---would be zero (because none of the point electron's mass is displaced from its center).  One might respond that in quantum physics we are forced to revise this definition of angular momentum and allow point particles to posses angular momentum.  The following sections show that there is no need to so radically revise our understanding of angular momentum.

The second obstacle is that an electron with the classical electron radius would have to spin superluminally to produce the correct magnetic moment.  Assuming the magnetic moment is generated by a spinning sphere of charge imposes essentially the same minimum radius for the electron as the first obstacle---the Compton radius.  The relation between velocity and magnetic moment $|\vec{m}|$ for a spherical shell of charge $-e$ is
\begin{equation}
|\vec{m}| = \frac{e R v}{3 c}\ 
\end{equation}
\citep[pg.\ 127]{rohrlich}.  Inserting $v=c$ and $|\vec{m}|=\frac{e \hbar}{2 m c}$ (the Bohr magneton) yields a radius of
\begin{equation}
R=\frac{3}{2}\frac{\hbar}{m c}\ .
\end{equation}

The third obstacle to regarding the electron as spinning is that its gyromagnetic ratio (the ratio of magnetic moment to angular momentum) differs from the simplest classical estimate (\citealp[problem 5.56]{griffiths}; \citealp[pg.\ 187]{jackson}):
\begin{equation}
\frac{|\vec{m}|}{|\vec{L}|}=\frac{e}{2 m c}\ .
\label{classicalGR}
\end{equation}
I stress that this is the \emph{simplest} estimate and not \emph{the} classical gyromagnetic ratio because its derivation requires two important assumptions beyond axial symmetry, each of which will be called into question later: (1) the mass $m$ and charge $-e$ are both distributed in the same way (i.e., mass density is proportional to charge density), and (2) the mass and charge rotate at the same rate.  With these assumptions in place, the derived gyromagnetic ratio is independent of the rate of rotation and the distribution of mass and charge.  The actual gyromagnetic ratio of the electron is twice this estimate,
\begin{equation}
\frac{|\vec{m}|}{|\vec{L}|}=\frac{e}{m c}\ ,
\label{quantumGR}
\end{equation}
as its angular momentum is $\frac{\hbar}{2}$ and its magnetic moment is the Bohr magneton, $\frac{e \hbar}{2 m c}$ (ignoring the anomalous magnetic moment).

The physicists who first proposed the idea of electron spin were aware of these obstacles.  Ralph Kronig was the first to propose a spinning electron to explain the fine structure of atomic line spectra (in 1925), but he did not publish his results because there were too many problems with his idea.  One of these problems was that the electron would have to rotate superluminally \citep[pg.\ 35]{tomonaga}.  Independently of Kronig, George Uhlenbeck and Samuel Goudsmit had the same idea.  Uhlenbeck spoke with Hendrik Lorentz about the proposal and Lorentz brought up the problem of superluminal rotation (among others).  After speaking with Lorentz, Uhlenbeck no longer wanted to publish.  But, it was too late.  His advisor, Paul Ehrenfest, had already sent the paper off.  Uhlenbeck recalls Ehrenfest attempting to reassure the pair by saying: ``You are both young enough to be able to afford a stupidity!'' (\citealp[pg.\ 47]{uhlenbeck}; see also \citealp{goudsmit}).  Uhlenbeck and Goudsmit were also aware of the gyromagnetic ratio problem, but they were not so troubled by it.  They understood that the classical calculation of the gyromagnetic ratio has assumptions that can be denied (e.g., the calculated gyromagnetic ratio would be different if the electron's mass were distributed evenly throughout the volume of a sphere and its charge were distributed over the surface; \citealp[pg.\ 47]{uhlenbeck}; \citealp[pg.\ 39]{pais1989}).

\section{The Electromagnetic Field}

\begin{figure}[htb]
\center{\includegraphics[width=8 cm]{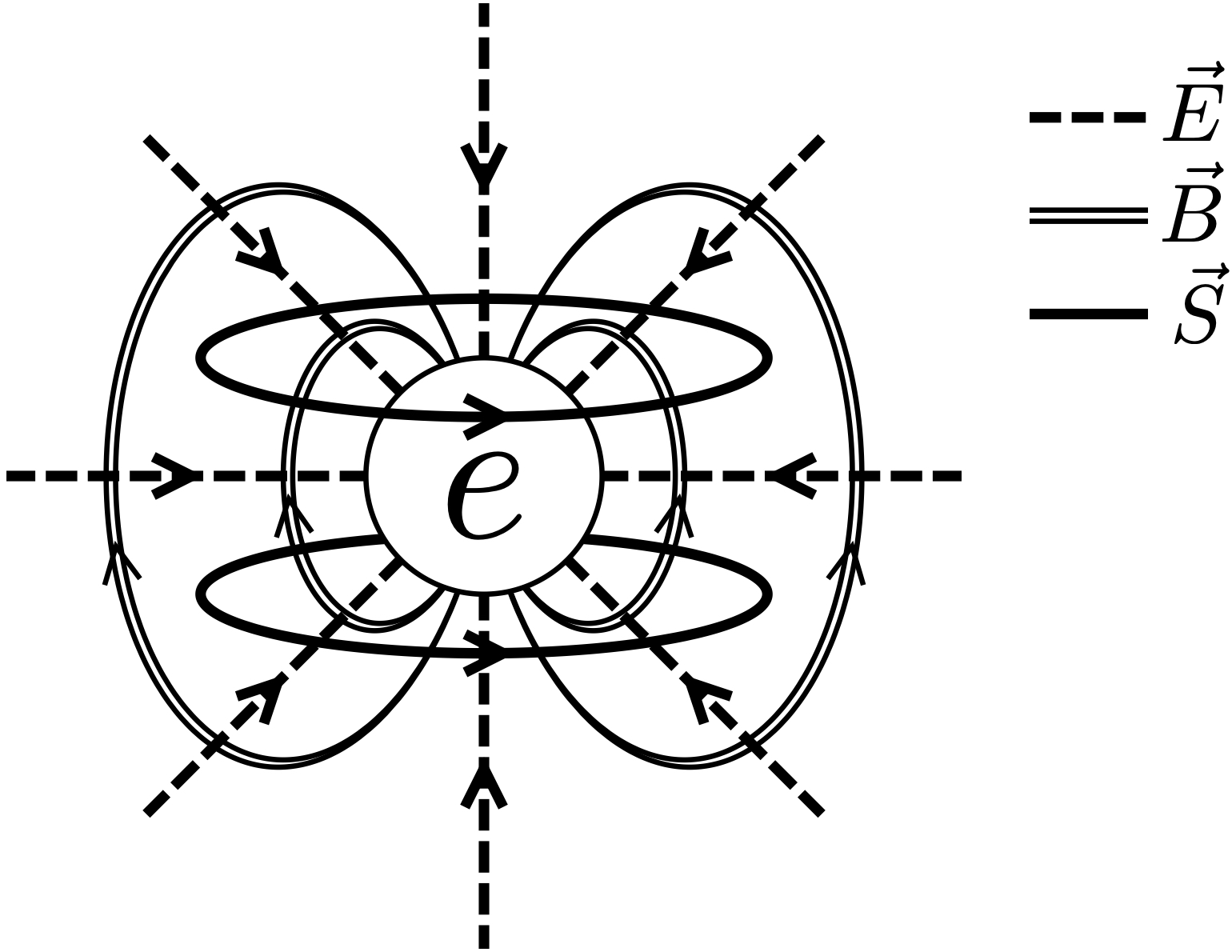}}
\caption{This figure depicts the electric and magnetic fields produced by the electron's charge and magnetic dipole moment.  Also shown is the Poynting vector $\vec{S}$ which indicates that the mass of the electromagnetic field rotates about the axis picked out by the electron's magnetic moment.}
  \label{chargeddipole}
\end{figure}

Before going on to model the electron using the Dirac field, it is worthwhile to consider how the above obstacles are altered by taking the mass of the electromagnetic field into account.  By the relativistic equivalence of mass and energy, the electromagnetic field has a relativistic mass density proportional to its energy density (see \citealp{lange}; \citealp{forcesonfields}).  In Gaussian units, the density of energy is
\begin{equation}
\rho_f^{\mathcal{E}}=\frac{1}{8 \pi}\left(|\vec{E}|^2+|\vec{B}|^2\right)\ ,
\label{energydensityfield}
\end{equation}
and the density of relativistic mass is
\begin{equation}
\rho_f=\frac{\rho_f^{\mathcal{E}}}{c^2}=\frac{1}{8 \pi c^2}\left(|\vec{E}|^2+|\vec{B}|^2\right) \ .
\label{massdensityfield}
\end{equation}
The $f$ subscript indicates that these are properties of the electromagnetic field and the $\mathcal{E}$ superscript distinguishes the energy density from the relativistic mass density.  The total mass of the electron is the sum of this electromagnetic mass plus any mass possessed by the electron itself.\footnote{I will use the phrase ``the electron itself'' to refer to the bare electron, distinct from the electromagnetic field that surrounds it.  This is to be contrasted with the dressed electron, which includes both the electron itself and its electromagnetic field.}  The mass of the electromagnetic field moves with a velocity that can be expressed in terms of the Poynting vector, $\vec{S}= \frac{c}{4\pi} \vec{E} \times \vec{B}$, which gives the energy flux density of the field.  The field velocity\footnote{This field velocity appears in \citet{poincare1900}; \citet[section 12.6.2]{holland}; \citet[box 8.3]{lange}; \citet{forcesonfields}.} can be found by dividing the energy flux density $\vec{S}$ by the energy density $\rho_f^{\mathcal{E}}$ or, equivalently, by dividing the momentum density of the field,
\begin{equation}
\vec{G}_f=\frac{\vec{S}}{c^2}= \frac{1}{4\pi c} \vec{E} \times \vec{B} \ ,
\label{momentumdensityfield}
\end{equation}
by its mass density \eqref{massdensityfield},
\begin{equation}
\vec{v}_f=\frac{\vec{G}_f}{\rho_f}=\frac{\vec{S}}{\rho_f^{\mathcal{E}}}\ .
\label{fieldvelocity}
\end{equation}
Looking at the field lines around a charged magnetic dipole, like the electron, it is clear from \eqref{momentumdensityfield} and \eqref{fieldvelocity} that the field mass circles the axis picked out by the dipole, as depicted in figure \ref{chargeddipole} (\citealp[chapter 27]{feynman2}; \citealp[chapter 8]{lange}).

The fact that some (or perhaps all) of the mass of the electron is located outside the bounds of the electron itself\footnote{Sometimes you see it said that a portion of the electron's mass is electromagnetic in origin, which seems to suggest that although this portion of mass originates in the energy of the electromagnetic field it is possessed by and located at the electron itself.  I have argued against such an understanding of electromagnetic mass in \citet{forcesonfields}.  The electromagnetic mass is located in the electromagnetic field.\label{whereisthemass}} and rotating appears to be helpful for addressing the first obstacle---getting a large angular momentum without moving superluminally is easier if the mass is more spread out.  Also, there is no danger of the mass in the electromagnetic field moving superluminally since the magnitude of the field velocity in \eqref{fieldvelocity} is maximized at $c$ when $\vec{E}$ is perpendicular to $\vec{B}$ and $|\vec{E}|=|\vec{B}|$.

We are now in a position to see where the classical electron radius comes from and to see why it is a completely unreasonable estimate to use in motivating the first obstacle (from section \ref{obstacles}).  Let's work up to that slowly.  First, note that the smaller the electron is, the greater the mass in the electric and magnetic fields surrounding the electron.  Keeping the total mass of the electron fixed, the smaller the electron is, the less mass it itself possesses.  If we imagine making the electron as small as possible,\footnote{If we were willing to make the mass of the electron itself negative, its radius could be even smaller \citep[pg.\ 214]{pearle}.} putting all of its mass in the electromagnetic field, we arrive at a radius for the electron that we can call the ``electromagnetic radius,'' on the order of\footnote{The exact number depends on the way the electron's charge is distributed and how that charge flows.} $10^{-12}$ cm (\citealp[pg.\ 47]{uhlenbeck}; \citealp[pg.\ 39]{pais1989}; \citealp[chapter 8]{macgregor}).  The classical electron radius was arrived at through similar reasoning applied before the electron's magnetic moment was discovered.  It was assumed that the electron's mass comes entirely from its electric field.  If we take the electron's charge to be distributed evenly over a spherical shell, the radius calculated in this way would be
\begin{align}
R&=\frac{e^2}{2 m c^2}\ ,
\label{electricsurfaceradius}
\end{align}
as the energy in the electric field is $\frac{e^2}{2 R}$.  Ignoring the prefactor (which is dependent on the way the charge is distributed), we get the classical electron radius,\footnote{See \citet[section 38-3]{feynman2}, \citet[section 6-1]{rohrlich}.}
\begin{equation}
R=\frac{e^2}{m c^2}\approx 2.82 \times 10^{-13} \mbox{ cm}\ .
\label{classicalelectronradius}
\end{equation}
This radius is an order of magnitude smaller than the electromagnetic radius because the amount of energy in the magnetic field is much greater than the amount in the electric field.  Neither of these radii should prompt worries of superluminal mass flow.  If the mass of the electron resides entirely in its electromagnetic field, then the electron itself is massless and energyless.  It doesn't matter how fast it's spinning since it itself won't have any angular momentum.  The angular momentum is entirely in the field and the mass of the field cannot move superluminally.

Understanding that the electromagnetic field possesses mass does little to alter the second obstacle.  Although the electron's mass bleeds into the field, its charge does not.

The third obstacle is complicated by the existence of mass in the electromagnetic field.  The simple calculation of the gyromagnetic ratio for a spinning charged body given above \eqref{classicalGR} was the ratio of the magnetic moment produced by a spinning body to the angular momentum of that body itself.  But, once we recognize that some of the mass we associate with that body is actually in its electromagnetic field, we must take the field's angular momentum into consideration when calculating the gyromagnetic ratio.  Here is one illustrative way of doing so.  The electric and magnetic fields around a charged magnetic dipole located at the origin are
\begin{align}
\vec{E}&=-e\frac{\vec{x}}{|\vec{x}|^3}
\nonumber
\\
\vec{B}&=\frac{3 (\vec{m}\cdot\vec{x})\vec{x}}{|\vec{x}|^5}-\frac{\vec{m}}{|\vec{x}|^3}\ .
\end{align}
If we assume the charge $-e$ to be uniformly distributed over a spherical shell of radius $R$, so that the above electric field is only present outside this radius, and also that the only contribution to the angular momentum comes from the fields outside this radius (because the entirety of the electron's mass resides in its electromagnetic field), we arrive at a gyromagnetic ratio of
\begin{equation}
\frac{|\vec{m}|}{|\vec{L}|}=\frac{3 R c}{2 e}\ .
\label{fieldGR}
\end{equation}
Unlike the simple calculation of the gyromagnetic ratio of an axially symmetric spinning charged body mentioned above \eqref{classicalGR}, this result is radius dependent.\footnote{I have only rarely seen the angular momentum of the electromagnetic field taken into account when calculating the gyromagnetic ratio of the electron (\citealp{corben1961, giulini2008}).}

We must input a radius for the electron if we are to compare \eqref{fieldGR} to \eqref{classicalGR} and \eqref{quantumGR}.  One option would be to use the classical electron radius \eqref{classicalelectronradius}.  However, we must be careful because the prefactors that were ignored in \eqref{classicalelectronradius} are important in \eqref{fieldGR}.  In the earlier derivation of the angular momentum of the field, it was assumed that the charge was distributed uniformly over the surface of the sphere as in \eqref{electricsurfaceradius}.  Continuing with that assumption and plugging \eqref{electricsurfaceradius} into \eqref{fieldGR} yields a gyromagnetic ratio of
\begin{equation}
\frac{|\vec{m}|}{|\vec{L}|}=\frac{3 e}{4 m c}\ ,
\label{firstGR}
\end{equation}
closer to \eqref{quantumGR} than \eqref{classicalGR}, but still incorrect.  The classical electron radius was calculated by ignoring the magnetic field of the electron.  Taking the magnetic field into consideration and using the electromagnetic radius instead of the classical radius would yield a gyromagnetic ratio which is much too large.  So, the assumption that the mass of the electron is entirely in the electromagnetic field leads to trouble.  Fortunately, it's not true.  In section \ref{restrictionsection} we'll see that the electron is large enough that the mass of the electromagnetic field surrounding the electron is only a small fraction of the electron's total mass.

\section{The Dirac Field}\label{diracfieldsection}

In the previous section we examined the flow of mass in the electromagnetic field surrounding the electron.  In this section we ignore the electromagnetic field and focus exclusively on the flow of mass and charge of the electron itself (assuming, contra the previous section, that little of the electron's mass is in the electromagnetic field).  We can understand this flow of mass and charge by using the Dirac field to represent the state of the electron.  In this section I make heavy use of the excellent account of spin given by \citet{ohanian}.

As was discussed in the introduction, the free Dirac equation,
\begin{equation}
i\hbar \frac{\partial \psi}{\partial t}=\left(\frac{\hbar c}{i}\gamma^0 \vec{\gamma}\cdot\vec{\nabla}+m\gamma^0 c^2\right)\psi
\ ,
\label{thediracequation}
\end{equation}
can either be viewed as part of a relativistic single particle quantum theory in which $\psi$ is a wave function (the quantum interpretation), or, as part of a relativistic field theory in which $\psi$ is a classical field (the classical interpretation).\footnote{As the Dirac field is sometimes interpreted as a wave function and sometimes as a classical field, one might naturally wonder if it is possible to interpret the electromagnetic field as a wave function instead of a classical field  (see \citealp{good1957}; \citealp{mignani1974}; \citealp{bialynicki1996}; \citealp{emasqp}).}  Here I adopt the second perspective and take $\psi$ to be a four-component complex-valued\footnote{Alternatively, the classical Dirac field is sometimes treated as Grassmann-valued (e.g., in textbook presentations of path integral methods for quantum field theory).  I discuss the relation between complex-valued classical Dirac field theory and Grassmann-valued classical Dirac field theory in \citet{positronpaper}.} classical field.  The classical Dirac field can be quantized, along with the classical electromagnetic field, to arrive at the quantum field theory of quantum electrodynamics.  In the context of quantum electrodynamics, the electron is described by a superposition of different states for the classical Dirac field (a wave functional).  In this paper, we will examine the classical field states that compose this superposition and see that our three obstacles can be overcome for each such classical state.  We will not need to go as far as quantizing the Dirac field.  At the level of physics under consideration here, there are just two interacting classical fields---the Dirac field and the electromagnetic field.\footnote{\citet[pg.\ 216--217]{weyl} explicitly considers and rejects the idea that the Dirac field should be treated as a classical field along the lines proposed here, comparing the idea to Schr\"{o}dinger's original pre-Born-rule interpretation of his eponymous equation where the amplitude-squared of the wave function is interpreted as a charge density.  It is true that before quantization the classical Dirac field does not provide an adequate theory of the electron (though such a theory works better than you might expect; see \citealp{crisp1969,jaynes1973,barut1988, barut1990}).  What matters for our purposes here is not the adequacy of classical Dirac field theory itself, but just the fact that it is this classical field theory which gets quantized to arrive at our best theory of the electron: quantum electrodynamics.  (It is worth noting that \citealp{weyl} later treats the Dirac field like a classical field when quantizing it; see \citealp[pg.\ 451]{pashby2012}.)}

Much like the electromagnetic field, the Dirac field carries energy and momentum.  The energy and momentum densities are given by:\footnote{These two densities are components of the symmetrized stress-energy tensor for the Dirac field (\citealp[section 20]{wentzel}; \citealp[appendix 7]{heitler}; \citealp[pg.\ 218--221]{weyl}).}$^,$\footnote{Here $\gamma^0$, $\vec{\gamma}$, and $\vec{\sigma}$ are four-dimensional matrices, related to the two-dimensional Pauli spin matrices $\vec{\sigma}_p$ by
\begin{equation}
\gamma^0=\left(\begin{matrix} I & 0 \\  0 & -I \end{matrix}\right)
\quad\quad
\vec{\gamma}=\left(\begin{matrix}0 & \vec{\sigma}_p \\  -\vec{\sigma}_p & 0 \end{matrix}\right)
\quad\quad
\vec{\sigma}=\left(\begin{matrix}\vec{\sigma}_p & 0 \\ 0 & \vec{\sigma}_p \end{matrix}\right)\ .
\label{matrixdefs}
\end{equation}}
\begin{align}
\rho_d^{\mathcal{E}}&=\frac{i \hbar}{2}\left(\psi^\dagger\frac{\partial \psi}{\partial t}-\frac{\partial \psi^\dagger}{\partial t}\psi\right)
\nonumber
\\
&=m c^2 \psi^\dagger\gamma^0\psi + \frac{\hbar c}{2i}\left[\psi^\dagger\gamma^0\vec{\gamma}\cdot\vec{\nabla}\psi-(\vec{\nabla} \psi^\dagger)\cdot\gamma^0\vec{\gamma}\psi\right] 
\label{diracenergydensity}
\\
\vec{G}_d&=\frac{\hbar}{2 i}\left[\psi^\dagger \vec{\nabla} \psi - (\vec{\nabla} \psi^\dagger)\psi \right]+\frac{\hbar}{4}\vec{\nabla}\times(\psi^\dagger \vec{\sigma} \psi)\ .
\label{diracmomentumdensity}
\end{align}
The $d$ subscript indicates that these are properties of the Dirac field.    The second term in the momentum density gives the contribution from spin (\citealp[pg.\ 181--182]{wentzel}, \citealp[pg.\ 168]{pauli}, \citealp[pg.\ 503]{ohanian}).  Because the spin contribution is a curl, it will not contribute to the total linear momentum of the electron.  When the momentum density in \eqref{diracmomentumdensity} is used to calculate the angular momentum of an electron, the first term yields the orbital angular momentum and the second yields the spin angular momentum.  The density of spin angular momentum derived from the second term in \eqref{diracmomentumdensity} is
\begin{equation}
\frac{\hbar}{2}\psi^\dagger \vec{\sigma} \psi\ .
\label{diracspinangularmomentumdensity}
\end{equation}
As we are here concerned with understanding spin, we will focus on states where the electron is at rest and the first term in \eqref{diracmomentumdensity} is everywhere zero.

Although I have not seen it done before, we can introduce a relativistic mass density and a velocity that describes the flow of mass in just the same way as was done for the electromagnetic field in the previous section,
\begin{align}
\rho_d&=\frac{\rho_d^{\mathcal{E}}}{c^2}=m \psi^\dagger\gamma^0\psi + \frac{\hbar}{2i c}\left[\psi^\dagger\gamma^0\vec{\gamma}\cdot\vec{\nabla}\psi-(\vec{\nabla} \psi^\dagger)\cdot\gamma^0\vec{\gamma}\psi\right] 
\label{diracmassdensity}
\\
\vec{v}_d&=\frac{\vec{G}_d}{\rho_d}\ .
\label{diracvelocity}
\end{align}
In contrast with the electromagnetic field, the Dirac field's energy density can be negative and thus its mass density can be negative as well.

In addition to the mass density and its flow, we can examine the charge density of the Dirac field and the flow of charge.  The charge density and charge current density are
\begin{align}
\rho^q_d&=-e \psi^\dagger \psi
\label{diracchargedensity}
\\
\vec{J}_d&=-e c \psi^\dagger \gamma^{0} \vec{\gamma} \psi
\label{diraccurrentdensity}\ .
\end{align}
If we were considering interaction with the electromagnetic field, these densities would act as source terms for Maxwell's equations.  From the charge and current densities, we can define the velocity of charge flow as
\begin{equation}
\vec{v}_d^{\:q}=\frac{\vec{J}_d}{\rho^q_d}=\frac{c \psi^\dagger \gamma^{0} \vec{\gamma} \psi}{\psi^\dagger \psi}\ .
\label{diracchargevelocity}
\end{equation}
From this definition, it follows that the charge velocity cannot exceed the speed of light (\citealp[section 2b]{takabayasi1957}; \citealp[section 10.4]{bohmhiley}; \citealp[section 12.2]{holland}).  Because of this light-speed limit, our second obstacle is automatically averted.  Superluminal charge flow is impossible.

The reason the gyromagnetic ratio of the electron differs from the simple classical estimate \eqref{classicalGR} by a factor of two can be explained straightforwardly in the context of Dirac field theory using the mass and charge velocities introduced above.\footnote{It is generally agreed that there exists some explanation of the gyromagnetic ratio in the context of the Dirac equation.  The task here is to better understand what sort of explanation is available (compare \citealp[pg.\ 504]{ohanian} and \citealp[section 1.4]{bjorkendrell}).}  In the simple estimate of the gyromagnetic ratio, we assumed that the mass and charge were rotating together at the same rate.  Actually, as we are about to see, the charge of the electron rotates twice as quickly as the mass.\footnote{How could charge move at a different velocity than mass?  Imagine you're describing a charged fluid flowing through pipes using certain mass and charge densities.  On closer inspection, the fluid turns out to be made of two kinds of particles---heavy neutral particles and light positively charged particles.  Sometimes the charged particles flow faster than the neutral ones and the velocity of charge flow is greater than the velocity of mass flow.  Sometimes the heavy particles flow faster than the light ones and the velocity of mass flow is greater than the velocity of charge flow.}  So, the magnetic moment is twice as large as you'd expect given the angular momentum.

This factor of two between the mass and charge velocity is a general feature of wave functions that describe an electron at rest.  But, to see how it arises it will be helpful to start with a particular illustrative example wave function.  Here is a simple instantaneous state of the Dirac field which we can use as a first approximation towards representing a single electron which is (at this moment) at rest with $z$-spin up:\footnote{This state is discussed in \citet[equation 12]{huang1952}; \citet[equation 3.32]{bjorkendrell}; \citet[equation 14]{ohanian}.}
\begin{equation}
\psi=\left(\frac{1}{\pi d^2}\right)^{3/4}e^{-|\vec{x}|^2/2d^2}\left(\begin{matrix} 1\\0\\0\\0 \end{matrix}\right)\ .
\label{ohanianstate}
\end{equation}
The mass and charge are both localized in a Gaussian wave packet of width $d$.  The reason for calling this a single electron state is that the integral of the charge density over all of space is $-e$.\footnote{See \citet[pg.\ 10]{takabayasi1957}.}

\begin{figure}[htb]
\center{\includegraphics[width=9.5 cm]{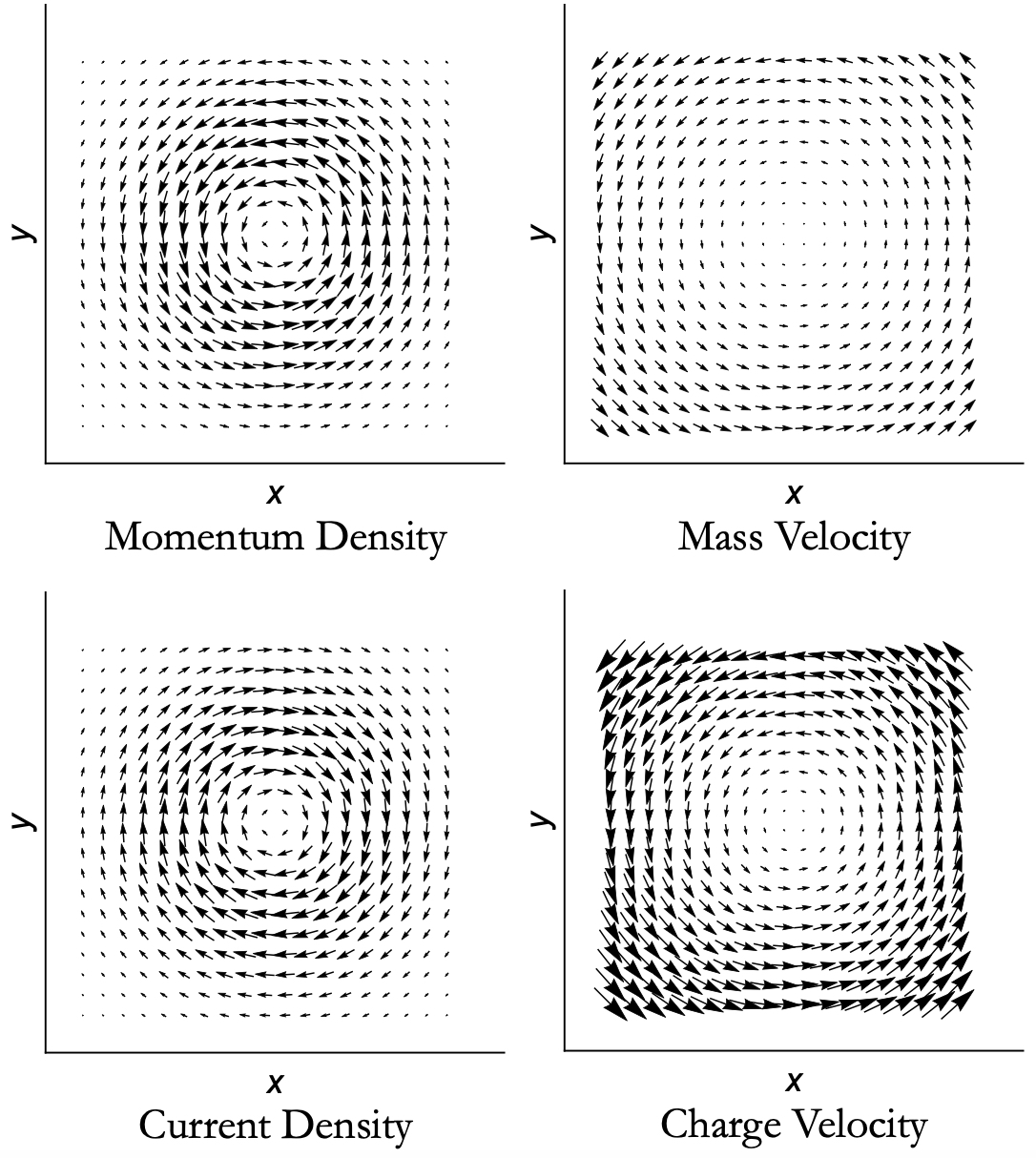}}
\caption{These plots depict the flow of mass and charge for the state of an electron at rest given in \eqref{ohanianstate}.  The first two plots give the momentum density \eqref{momex} and mass velocity \eqref{massvel}.  The second two plots give the magnetization current density \eqref{magcurrent} and the corresponding contribution to the charge velocity \eqref{chargevel} (for the corrected state \eqref{approxstate}, these plots give the total charge current density and charge velocity).  The two velocity plots use the same scale to highlight that the charge velocity is twice the mass velocity.}
\label{flows}
\end{figure}

The momentum density for this state is
\begin{equation}
\vec{G}_d=-\frac{\hbar}{2}\left(\frac{1}{\pi d^2}\right)^{3/2}e^{-|\vec{x}|^2/d^2}\ \frac{\vec{x}\times\hat{z}}{d^2}\ ,
\label{momex}
\end{equation}
calculated via \eqref{diracmomentumdensity} where only the second term is non-zero.  From this expression, it is clear that mass and energy are flowing around the $z$-axis (see figure \ref{flows}).  The mass velocity for this state can be calculated by dividing this momentum density by the mass density, as in \eqref{diracvelocity},
\begin{equation}
\vec{v}_d = - \frac{\hbar}{2 m}\frac{\vec{x}\times\hat{z}}{d^2}= \frac{\hbar r}{2 m d^2}\hat{\theta}\ .
\label{massvel}
\end{equation}
The second expression gives the velocity in cylindrical coordinates.  This equation shows that the mass flows everywhere about the $z$-axis at constant angular velocity.  The electron's mass appears\footnote{I use the qualification ``appears'' because, as will be explained shortly, \eqref{ohanianstate} is not an entirely satisfactory approximation to the state of an electron.} to rotate like a solid object.

To calculate the velocity at which charge flows, it is useful to first expand the current density using the free Dirac equation as follows\footnote{This expansion appears in \citet{gordon1928}; \citet[pg.\ 321--322]{frenkel}; \citet[pg.\ 479]{huang1952}; \citet[pg.\ 504]{ohanian}.}
 \begin{equation}
 -e c \psi^\dagger \gamma^{0} \vec{\gamma} \psi = \underbrace{\strut \frac{i e\hbar}{2 m}\left\{\psi^\dagger \gamma^0 \vec{\nabla} \psi - (\vec{\nabla} \psi^\dagger) \gamma^0\psi\right\}}_{\text{\large{\ding{172}}}}\underbrace{\strut  - \frac{e\hbar}{2 m} \vec{\nabla}\times(\psi^\dagger \gamma^0 \vec{\sigma}\psi)}_{\text{\large{\ding{173}}}}\underbrace{\strut  +\frac{i e\hbar}{2 m c}\frac{\partial}{\partial t}(\psi^\dagger \vec{\gamma} \psi)}_{\text{\large{\ding{174}}}}\ .
\label{currentexpansion}
 \end{equation}
The three terms in the expansion are the convection current density, the magnetization current density, and the polarization current density.  As was the case for the momentum density \eqref{diracmomentumdensity}, the first term is zero for an electron at rest.  The second two terms give the contribution to the charge current from spin.  For the moment, let us focus on the magnetization current density.  The magnetization current density in \eqref{currentexpansion} corresponds to a magnetic moment density of
\begin{equation}
- \frac{e\hbar}{2 m c} \psi^\dagger \gamma^0 \vec{\sigma}\psi\ ,
\label{magneticmomentdensity}
\end{equation}
where the prefactor is the Bohr magneton.\footnote{See \citet[section 5.6]{jackson}; \citet[pg. 504]{ohanian}.}  The ratio of the magnitude of this magnetic moment density to the magnitude of the angular moment density in \eqref{diracspinangularmomentumdensity} for the state in \eqref{ohanianstate} is $\frac{e}{m c}$, the correct gyromagnetic ratio for the electron \eqref{quantumGR}.  The magnetization current density,
\begin{equation}
\frac{e \hbar}{m}\left(\frac{1}{\pi d^2}\right)^{3/2}e^{-|\vec{x}|^2/d^2}\ \frac{\vec{x}\times\hat{z}}{d^2}\ ,
\label{magcurrent}
\end{equation}
makes a contribution to the velocity of charge flow, calculated via \eqref{diracchargevelocity}, of
\begin{equation}
-\frac{\hbar}{m}\frac{\vec{x}\times\hat{z}}{d^2}=\frac{\hbar r}{m d^2} \hat{\theta}\ .
\label{chargevel}
\end{equation}
The contribution to the velocity of charge flow which determines the electron's magnetic moment \eqref{chargevel} is twice the velocity of mass flow which determines the electron's angular momentum \eqref{massvel}.

The factor of two between these velocities is not a peculiar feature of the chosen state, but will hold for any electron state in the non-relativistic limit.  In general, the contribution to the moment density $\vec{G}_d$ from spin is $\frac{\hbar}{4}\vec{\nabla}\times(\psi^\dagger \vec{\sigma} \psi)$---the second term in \eqref{diracmomentumdensity}.  In the non-relativistic limit, the relativistic mass density is approximately $m \psi^\dagger\gamma^0\psi$---the first term in \eqref{diracmassdensity}.  Dividing these, as in \eqref{diracvelocity}, gives a contribution to the velocity of mass flow from spin of
\begin{equation}
\frac{\hbar}{4 m} \frac{\vec{\nabla}\times(\psi^\dagger \vec{\sigma} \psi)}{\psi^\dagger\gamma^0\psi}\ .
\label{massvelgen}
\end{equation}
The velocity associated with the electron's spin magnetic moment can be derived from the magnetization current density---the second term in \eqref{currentexpansion}.  Dividing the magnetization current density by the charge density \eqref{diracchargedensity}, as in \eqref{diracchargevelocity}, yields a contribution to the charge velocity of
\begin{equation}
\frac{\hbar}{2 m} \frac{\vec{\nabla}\times(\psi^\dagger \gamma^0\vec{\sigma} \psi)}{\psi^\dagger\psi}\ .
\label{chargevelgen}
\end{equation}
It is clear that \eqref{chargevelgen} is twice \eqref{massvelgen} (up to factors of $\gamma^0$ which we will return to later).

Addressing our obstacles in the context of the Dirac equation has enabled significant progress, but there remain three serious shortcomings to the account given thus far (which will be resolved in the following section).  First, we have only been able to say (somewhat awkwardly) that a certain contribution to the velocity of charge flow is twice the velocity of mass flow and not that the actual velocity of charge flow is twice the velocity of mass flow.  In fact, it is easy to see that the velocity of charge flow is zero for the state in \eqref{ohanianstate} as the charge current density calculated from \eqref{diraccurrentdensity} is clearly zero.  The first term in the current expansion \eqref{currentexpansion} is also zero.  Thus, the third term in \eqref{currentexpansion} (the polarization current density) must exactly cancel the second (the magnetization current density).  Because of this cancellation, no magnetic field is being produced by an electron in this state.  If we want to account for the magnetic field around an electron at rest, we need the electron's charge to actually rotate.

Second, the velocities in \eqref{massvel} and \eqref{chargevel} are unbounded, becoming superluminal as $r$ becomes very large.  The fact that \eqref{chargevel} becomes infinite is not so troubling because, as was just discussed, it is cancelled by the contribution to the charge velocity from the polarization current.  Also, as was mentioned earlier, it can be shown in general that the charge velocity cannot exceed $c$.  The fact that \eqref{massvel} becomes superluminal is a real problem.

Third, there are problems that arise if the electron is too small and as of yet we have no reason to think it's large enough to avoid these problems.  If the electron is too small, we face our first two obstacles concerning superluminal rotation.  Also, if the electron is too small we will not be able to ignore the mass in the electromagnetic field when calculating the gyromagnetic ratio (as was done in this section but not the last).  Looking at \eqref{ohanianstate} it appears that the size of the electron is an entirely contingent matter depending on the state of the Dirac field (by decreasing $d$, the electron can be made arbitrarily small).

\section{Restriction}\label{restrictionsection}

The three problems raised at the end of the previous section can be resolved by restricting the allowed states of the Dirac field to those formed by superposing positive frequency modes.  Such a restriction can be motivated by the fact that, in quantum field theory, the positive frequency modes of the Dirac field are associated with electrons and the negative frequency modes with positrons.

In this section we will continue to focus our attention on the free Dirac equation, putting aside issues of self-interaction---the electron is treated as blind to the electromagnetic field it generates---but still confronting issues of self-energy---the energies of the Dirac and electromagnetic fields are both taken into account.

The free Dirac equation admits of plane wave solutions with definite\footnote{On a classical interpretation of the Dirac field, by saying the momentum is ``definite'' I mean that the momentum density \eqref{diracmomentumdensity} is uniform.} momentum $\vec{p}$ and time dependence given by either $e^{-i \mathcal{E}_{\vec{p}} t /\hbar}$ (positive frequency) or $e^{i \mathcal{E}_{\vec{p}} t/\hbar}$ (negative frequency), where $\mathcal{E}_{\vec{p}}$ is the energy associated with that momentum, $\mathcal{E}_{\vec{p}}=\sqrt{|\vec{p}|^2 c^2 + m^2 c^4}$ \citep[chapter 3]{bjorkendrell}.  From \eqref{diracenergydensity}, it is clear that the positive frequency plane waves have uniform positive energy density and the negative frequency plane waves have uniform negative energy density.

In textbook presentations of the quantization of the classical Dirac field (such as \citealp[section 3.5]{peskinschroeder}), one starts by expanding the Dirac field in terms of positive and negative frequency modes.  Quantizing this theory in a straightforward way, one pairs creation operators with each of these modes and sees that the operators paired with the positive frequency modes create particles with negative charge and positive energy whereas the operators paired with the negative frequency modes create particles with negative energy and negative charge.  Seeking to avoid negative energies, one then redefines the operators for charge and energy so that the operators associated with negative frequency modes can be reinterpreted as annihilation operators for particles with positive energy and negative charge---positrons.  Bringing the lesson that negative frequency modes will ultimately be associated with positrons back to our classical Dirac field theory,\footnote{In \citet{positronpaper}, I describe in more detail the procedure of field quantization and the lessons one ought to draw from it for classical Dirac field theory.} we ought to revise our understanding of the electron within classical Dirac field theory.  In representing the electron, we will forbid any state of the Dirac field which has a Fourier decomposition that includes negative frequency modes and permit only states that are formed entirely from positive frequency modes (reserving the negative frequency modes for the representation of positrons).  We will thus focus on a restricted version of classical Dirac field theory where the negative frequency modes are removed.  (As we are not considering interactions, if negative frequency modes are absent at one time they will be absent always.)  Before continuing on to analyze the electron within this restricted classical Dirac field theory, let us take a brief detour to examine the way negative frequency modes were originally handled by Dirac.

On a quantum interpretation of the Dirac equation where the Dirac field is viewed as a wave function, one would say that the negative frequency plane wave solutions are energy eigenstates with negative eigenvalues.  The existence of such negative energy states proved both a blessing and a curse for early applications of the Dirac equation.  To retain the blessing while dispelling the curse, Dirac proposed his hole theory according to which the negative energy states are filled \citep{dirac1930theory}.\footnote{See \citet{saunders1991, pashby2012}.}  By Pauli exclusion, any additional electrons must sit atop this ``Dirac sea.''  The filled sea is taken to set the zero level for energy and charge.  If any electron is excited out of the sea, the hole it leaves behind acts like a particle with equal mass and opposite charge to the electron---a positron.  In this ``hole theory,'' the positive frequency modes (which are by default empty) are used to describe ordinary positive energy electrons and the negative frequency modes (which are by default filled) are used to describe positrons.\footnote{Some authors use the idea of Dirac sea in presenting the quantization of the Dirac field and others emphatically renounce it (compare \citealp[section 8a]{schweberQFT}; \citealp[section 13.4]{bjorkendrellfields}; \citealp{hatfield} to \citealp[chapter 2]{duncan}; \citealp[pg.\ 142]{schwartz}).}
 
As was first examined by \citet{weisskopf1934a, weisskopf1934b, weisskopf1939}, the electromagnetic energy divergence---which arises because the amount of energy in an electron's electromagnetic field goes rapidly to infinity as its radius is decreased---is tamed in the context of hole theory (\citealp[section 2.5.3]{schweber1994}).  Weisskopf's handling of this divergence has been incorporated into the modern understanding of mass renormalization within quantum electrodynamics.\footnote{It is cited in relation to a modern understanding based on Feynman diagrams by \citet[pg.\ 513]{schweberQFT}; \citet[pg.\ 165]{bjorkendrell}; \citet[section II.D.2]{weisskopf1986}.}  The crucial insight from Weisskopf's analysis for our task at hand is expressed well by \citet[pg.\ 299]{heitler}.  He writes that the taming of the electromagnetic self-energy divergence for an electron at rest ``is a consequence of the hole theory and the Pauli [exclusion] principle'':
\begin{quote}
``Consider an electron represented by a very small wave packet in coordinate space.  In momentum space this would be represented by a distribution including negative energy states.  The latter, however, are filled with vacuum electrons.  Consequently, the negative energy contributions to the wave function must be eliminated and the electron cannot be a wave packet of infinitely small size but must have a finite extension (of the order $\frac{\hbar}{mc}$ [the Compton radius], as one easily finds).  Consequently the static self-energy will be diminished also.''\footnote{The fact that wave packets composed of positive energy modes have a minimum size is also discussed in \citet{newton1949}; \citet[pg.\ 39]{bjorkendrell}; \citet{chuu2007}.\label{minsizesources}}
\end{quote}

Although Heitler's explanation is given from within the quantum interpretation of the Dirac field as wave function, it contains lessons that carry over to our classical interpretation.  There is a limit on the minimum size wave packet that one can construct from the positive frequency modes of the Dirac field that are available in restricted Dirac field theory.  The mass and charge of the electron thus cannot be confined to an arbitrarily small volume.  Because the charge of the electron is spread over such a large packet, the electromagnetic contribution to the energy (and mass) of an electron at rest is small and can be ignored when calculating the gyromagnetic ratio to a first approximation (as was done in the previous section).  In section \ref{obstacles} we saw that in order to avoid the superluminal rotation speeds forced upon us in our first two obstacles, the electron must be at least as large as the Compton radius.  Restricting to positive frequency modes delivers that minimum size.

Let us return to the instantaneous electron state in \eqref{ohanianstate}.  In restricted Dirac field theory, this state is forbidden as it includes both positive and negative frequency modes.  To find a similar state that is allowed, we can simply delete the negative frequency modes from the Fourier decomposition of \eqref{ohanianstate}.\footnote{This Fourier decomposition is given in \citet[section 3.3]{bjorkendrell}.}  This yields\footnote{As the negative frequency modes have simply been deleted, this new state is not normalized.  In our classical terms, this means that the integral of the charge density over all space will not be $-e$.  In the non-relativistic limit, the total charge will be close to $-e$.}
\begin{equation}
\psi=\frac{1}{2}\left(\frac{d^2}{\pi \hbar^2}\right)^{3/4}\left(\frac{1}{2 \pi \hbar}\right)^{3/2}\int d^3 p \left(1+\frac{m c^2}{\mathcal{E}_{\vec{p}}}\right)e^{-\frac{|\vec{p}|^2 d^2}{2 \hbar^2}+\frac{i}{\hbar}\vec{p}\cdot\vec{x}}\left(\begin{matrix}
1\\
0\\
\frac{p_z c}{\mathcal{E}_{\vec{p}} + m c^2}\vspace*{4 pt}\\
\frac{(p_x+ i p_y) c}{\mathcal{E}_{\vec{p}} + m c^2}
\end{matrix}\right)\ .
\label{truestate}
\end{equation}
We can approximate this state in the non-relativistic limit by computing these integrals assuming that $d \gg \frac{\hbar}{m c}$ so the momentum space Gaussian in the integrand suppresses modes where $|\vec{p}|^2$ is not $\ll m^2 c^2$,\footnote{The same approximation can be arrived at by trusting only the first two components of \eqref{ohanianstate} and using the positive frequency non-relativistic limit of the Dirac equation in \citet[equation 1.31]{bjorkendrell} to calculate the other two.}
\begin{equation}
\psi=\left(\frac{1}{\pi d^2}\right)^{3/4}e^{-|\vec{x}|^2/2d^2}\left(\begin{matrix}
1\\
0\\
\frac{\hbar}{2 m c d^2}i z\vspace*{4 pt}\\
\frac{\hbar}{2 m c d^2}(i x-y)
\end{matrix}\right)\ .
\label{approxstate}
\end{equation}

The total current density for this state \eqref{approxstate}, calculated via \eqref{diraccurrentdensity}, is equal to the previous magnetization current density \eqref{magcurrent}.  Dividing this by the charge density for \eqref{approxstate} gives a charge velocity of
\begin{equation}
\vec{v}_d^{\:q}=\frac{\frac{-\hbar}{m d^2}\vec{x}\times\hat{z}}{1+\frac{\hbar^2}{m^2 c^2 d^4}|\vec{x}|^2}\ .
\label{chargevel2}
\end{equation}
This limits to \eqref{chargevel} for $d \gg \frac{\hbar}{m c}$.  Unlike \eqref{chargevel}, this is the actual charge velocity and not merely a contribution to it.  The charge is really moving.  The velocity in \eqref{chargevel2} is bounded and will not exceed the speed of light---as must be the case since the definition of the charge velocity \eqref{diracchargevelocity} ensures that it cannot be superluminal.  For $d \gg \frac{\hbar}{m c}$, the mass velocity derived from \eqref{approxstate} using \eqref{diracvelocity} will be as it was before \eqref{massvel}.  Thus, the charge rotates twice as fast as the mass.

This factor of two between mass and charge velocity is a general feature of states that describe electrons at rest in restricted Dirac field theory.  What prevented us from reaching this conclusion in the previous section was that we had no reason to suppose the magnetization current density would be the dominant contribution to the total current density \eqref{currentexpansion}.  The polarization current density could be significant as well.  By restricting ourselves to superpositions of positive frequency modes, we have guaranteed that the polarization current density is small.  To see why this is so, consider an arbitrary state of the Dirac field at $t=0$, $\psi(0)$.  This state can be written as the sum of a superposition of positive frequency modes, $\psi_+$, and a superposition of negative frequency modes, $\psi_-$.  In the non-relativistic limit, the time dependence of this state is given by
\begin{equation}
\psi(t)=e^{(i m c^2 / \hbar) t}\psi_++e^{-(i m c^2 / \hbar) t}\psi_-\ .
\end{equation}
The polarization current density for this state is
\begin{equation}
\frac{i e\hbar}{2 m c}\frac{\partial}{\partial t}\big(\psi_+^\dagger \vec{\gamma} \psi_++e^{-2(i m c^2 / \hbar) t}\psi_+^\dagger \vec{\gamma} \psi_-+e^{2(i m c^2 / \hbar) t}\psi_-^\dagger \vec{\gamma} \psi_++\psi_-^\dagger \vec{\gamma} \psi_-\big)\ .
\end{equation}
If we forbid negative frequency modes, the cross terms are absent and the time derivative yields zero.  Thus, in the non-relativistic limit the polarization current density is negligible.

In the previous section we were able to derive the factor of two between charge velocity and mass velocity only up to factors of $\gamma^0$.  The reason these factors can be ignored is that $\gamma^0$ simply flips the sign of the third and fourth components of $\psi$ and these components---for a state composed of positive frequency modes in the non-relativistic limit---are much smaller than the first and second components.

At this point let us reflect on the role that the non-relativistic limit has played in the preceding analysis.  This limit is not part of the general response to our first two obstacles.  The fact that there is a minimum size for wave packets formed from positive frequency modes is not dependent on this limit, nor is the light-speed cap on charge velocity.  The non-relativistic limit is, however, essential in explaining the electron's gyromagnetic ratio.  The reason for this is that the gyromagnetic ratio we seek to account for only holds in the non-relativistic limit.\footnote{Standard explanations of the factor of two in the gyromagnetic ratio of the electron using the Dirac equation appeal to the non-relativistic limit (\citealp[section 1.4]{bjorkendrell}).}  Beyond this limit, the relationship between angular momentum and magnetic moment is more complex.  In quantum mechanical terms, the relationship is given by the claim that the spin magnetic moment operator, $- \frac{e\hbar}{2 m c} \gamma^0 \vec{\sigma}$, is $-\frac{e\gamma^0}{m c}$ times the spin angular momentum operator, $\frac{\hbar}{2}\vec{\sigma}$ (\citealp[pg.\ 504]{ohanian}; \citealp[pg.\ 323]{frenkel}).  Expressed in terms of local expectation values, the local ratio of spin magnetic momentum to angular momentum is the ratio of $- \frac{e\hbar}{2 m c} \psi^\dagger \gamma^0 \vec{\sigma}\psi$ to $\frac{\hbar}{2}\psi^\dagger \vec{\sigma} \psi$.  In our classical field terminology, this is understood as the ratio of the spin magnetic moment density \eqref{magneticmomentdensity} to the spin angular momentum density \eqref{diracspinangularmomentumdensity}.

\section{Other Accounts of Spin}

In the introduction, I distinguished between field and particle approaches to quantum field theory.  In the field approach, one starts from classical Dirac field theory and then quantizes.  In the particle approach, one starts from a single electron relativistic quantum theory and then extends to multiple particles.  Although I am not aware of other authors who explicitly argue that the electron really rotates within the field approach, there have been a number of attempts to somehow understand the electron's angular momentum and magnetic moment as resulting from the motion of the electron's mass and charge within the particle approach---where one sees $\psi$ as the electron's quantum wave function.\footnote{\citet{ohanian} calls $\psi$ a ``wave field,'' which is confusingly ambiguous between quantum wave function and classical field.  Because of the way he uses quantum language, I have classified him as adopting a particle approach where $\psi$ is seen as a quantum wave function.}  In this section, I will briefly compare the analysis presented here to a few of these accounts of spin.  To organize the discussion, I will sort the accounts into two classes.  First, there are those who---despite sometimes using quantum language---treat $\psi$ as broadly similar to a classical field, putting aside the probabilistic nature of this wave function and understanding the mass and charge of the electron to be spread out and executing some rotational motion.  Second, there are those who emphasize the probabilistic nature of $\psi$ and think of the electron's mass and charge as located at a point.  On such an account, the angular momentum and magnetic moment of the electron are not explained by a spinning motion (as the electron is point size) but instead by a rapid circulation of the electron within its wave function.

Let us begin with the first class.  The most similar account of spin to the one proposed here is that of \citet{ohanian}.  He describes the flow of energy and charge in the electron's wave function just as I describe the flow of energy and charge in the Dirac field in section \ref{diracfieldsection}---though he introduces neither a velocity of mass flow \eqref{diracvelocity} nor a velocity of charge flow \eqref{diracchargevelocity}.  Ohanian explains that the angular momentum and magnetic moment of the electron are not ``internal'' and ``irreducible,'' but instead result from these flows of energy and charge.  Ohanian does not directly address the three obstacles raised in section \ref{obstacles} and does not make the moves in section \ref{restrictionsection} that are necessary to surmount them.  \citet{chuu2007} also give an account of spin where the electron's mass and charge are understood to be spread over the wave function and rotating.  However, their description of this flow is somewhat different from that of section \ref{diracfieldsection} as they use the same current for the flows of both mass and charge.  Chuu \textit{et al.\ }note that a wave packet formed from positive energy modes has a minimum size large enough to avoid the first obstacle (as was important in section \ref{restrictionsection}).  This minimum size can also be used to address the second obstacle, though they do not mention that explicitly.  They do not give a response to the third obstacle.

To understand the accounts that fall within the second class, let us start by examining the flow of probability.  One can introduce densities of probability and probability current for the electron's wave function that are proportional to the densities of charge and charge current from section \ref{diracfieldsection}, \eqref{diracchargedensity} and \eqref{diraccurrentdensity}.  The probability density is $\psi^\dagger \psi$ and the probability current density is $c\psi^\dagger\gamma^{0} \vec{\gamma}\psi$.  This probability current density is the local expectation value of the ``velocity operator'' $c\gamma^{0} \vec{\gamma}$ (or, equivalently, $c\vec{\alpha}$).\footnote{This velocity operator is presented in \citet[sections 31 and 32]{frenkel}; \citet[section 69]{dirac}; \citet[pg. 920--922]{messiah1962}; \citet[pg.\ 11]{bjorkendrell}.}  I did not introduce these densities earlier because I was treating $\psi$ as a classical field and thus all of this quantum talk about the flow of probability would have been inappropriate.  The classical Dirac field has a mass density and a charge density, but no probability density.  Now that we are treating $\psi$ as a quantum wave function, it is important to introduce a probability density and probability current density.

Applying a Bohmian interpretation of quantum mechanics to this single particle relativistic quantum theory, one can posit the existence of a point electron particle that is separate from the Dirac wave function and guided by it.  Dividing the probability current density by the probability density yields a velocity for this particle, $\frac{c\psi^\dagger\gamma^{0} \vec{\gamma}\psi}{\psi^\dagger\psi}$, equal to the velocity of charge flow \eqref{diracchargevelocity} introduced earlier and thus also capped at $c$ (\citealp{bohm1953comments}; \citealp[section 10.4]{bohmhiley}; \citealp[equation 12.2.10]{holland}).  Taking the electron to be a point particle moving with this velocity leads to the possibility of understanding the electron's angular momentum and magnetic moment as generated by the electron's motion within its wave function.  \citet[pg.\ 218]{bohmhiley} argue that: ``in the Dirac theory, the magnetic moment usually attributed to the `spin' can actually be attributed to a circulating movement of a point particle, and not that of an extended spinning object.''

\citet{huang1952} also thinks of the electron as a point particle whose circulating motion gives rise to the observed angular momentum and magnetic moment, but he has different ideas about how it moves.  He writes:
\begin{quote}
``... in the Dirac theory [velocity] is represented by the operator [$c\gamma^{0} \vec{\gamma}$], whose components can only have eigenvalues of $\pm c$, where $c$ is the velocity of light.  This means that while the average velocity of the electron is less than $c$, its instantaneous velocity is always $\pm c$.  We infer from this that the motion of the electron consists of a highly oscillatory component, superimposing on the average motion.  Schr\"{o}dinger called this oscillatory motion `zitterbewegung,' and showed that the amplitude of this oscillation is of the order of the Compton wavelength of the electron.''\footnote{This conjectured zitterbewegung (``trembling motion'') of the electron is discussed in \citet[section 69]{dirac}; \citet[pg.\ 38]{bjorkendrell}.  \citet{barutzanghi} present a way of using zitterbewegung to understand electron spin that is different from the proposals of Huang and Hestenes.}
\end{quote}
The existence of a Bohmian theory in which the electron's velocity can be less than the speed of light shows that we are not forced to regard the electron as always traveling at the speed of light.  But, we can see where that thought leads.  Unlike \citet{bohmhiley}, Huang does not give general equations for calculating the electron's motion.  In the article, he considers the example state in \eqref{ohanianstate} and shows that it can be written as a superposition of states in which the expectation value of position executes circular motion (though the expectation value of position for the actual state does not move).  He takes this observation and others to suggest that in this state the electron circles the $z$-axis.

Making Huang's picture precise, Hestenes has proposed equations of motion for a point electron according to which the electron always moves at the speed of light (see \citealp{gull1993}; \citealp{hestenes2008, hestenes2010} and references therein).  The angular momentum and magnetic moment arise from a circulating motion of the electron (depicted in figure 1 of \citealp{hestenes2008}).  According to \citet[pg.\ 2]{hestenes2010}, it may be possible to view the equations he proposes as ``formulating fundamental properties of the electron that are manifested in the Dirac equation in some kind of average form.''  Hestenes recognizes that his equations represent a departure from standard physics and has considered ways of empirically observing the predicted deviations (\citealp{hestenes2008}; \citealp[sections 9.1 and 11]{hestenes2010}).  He also acknowledges that there is work to be done in reconciling his novel approach with quantum field theory as we know it \citep[pg.\ 53]{hestenes2010}.  This kind of modificatory program is quite different from the account of spin provided here, where I've sought to show that our existing equations, properly interpreted, describe a spinning electron.

\section{Conclusion}

The consensus about electron spin, which emerged long ago, is that the electron somehow acts like a spinning object without actually spinning.  As \citet[pg.\ 514]{rojansky} puts it in his textbook on quantum mechanics, after discussing angular momentum and magnetic moment in the context of the Dirac equation,
\begin{quote}
``In short, \emph{Dirac's equation automatically endows the electron with the properties that account for the phenomena previously ascribed to a hypothetical spinning motion of the electron.}'' (original italics)
\end{quote}
In this paper I have argued for a different interpretation. The Dirac equation does not somehow manage to account for these properties without positing a spinning electron.  Instead, it explains just how the electron spins.

The obstacles to regarding the electron as spinning presented in the introduction were addressed as follows:  Old estimates of the size of the electron made under the assumption that the electron's mass is primarily electromagnetic suggest that the electron would have to rotate superluminally in order to have the right angular momentum and magnetic moment.  Actually, if the electron's mass is primarily electromagnetic we should focus on the rotation of the electromagnetic field's mass in calculating the electron's angular momentum and this mass cannot move superluminally.  Also, the electron's mass is not primarily electromagnetic.  When we move to better estimates of the electron's size---using the Dirac field to represent the state of the electron---we see that its minimum size is large enough that there is no need for superluminal rotation.  Further, the definition of charge velocity for the Dirac field guarantees that the electron's charge will not move superluminally.  The other obstacle was the fact that the electron's gyromagnetic ratio differs from the simplest classical estimate by a factor of two.  On the account given here, this factor does not arise from some novel quantum revision to the basic physical principles defining angular momentum and magnetic moment, but is instead attributed to a false assumption in the simple classical estimate---the electron's mass and charge do not rotate at the same rate.\\

\textbf{Acknowledgments}
Thank you to Adam Becker, Dirk-Andr\'{e} Deckert, Maaneli Derakhshani, John McGreevy, Lukas Nickel, Hans Ohanian, Laura Ruetsche, Roderich Tumulka, David Wallace, and anonymous referees for helpful feedback and discussion.  This project was supported in part by funding from the President's Research Fellowships in the Humanities, University of California (for research conducted while at the University of California, San Diego).

\section*{Note on Further Developments}

The text above is as in the 2019 published version, except that equation 2 has been corrected and references that were not yet published have been filled in.  Here I am adding a note in July, 2024.

Although I still think that the picture of electron spin as true rotation presented here is largely correct, there is a piece of the story that turned out to be incorrect and there are lingering puzzles that have yet to be fully resolved.  In section \ref{restrictionsection}, I quoted \citet{heitler} (and cited other sources in footnote \ref{minsizesources}) for the claim that there is a minimum size, on the order of the Compton radius, for electron wave packets constructed from positive frequency modes.  This minimum size was then used to explain why there is no need for superluminal energy or charge flow to account for the angular momentum and magnetic moment of the electron within classical Dirac field theory (addressing the first and second obstacles from section \ref{obstacles}, supplemented by the point under \eqref{diracchargevelocity} that the charge velocity for the Dirac field cannot exceed the speed of light).  It turns out that Heitler's claim of a minimum size can be disproven by explicit counterexamples.  It is possible to have arbitrarily compact electron wave packets formed entirely from positive frequency modes.  There is no minimum size.

Where does this leave us with regard to the first two obstacles from section \ref{obstacles}?  Here is how I summarize the situation in a 2022 article (omitting the footnotes that appear there), mentioning two articles in {\em Physical Review A} that further explore these issues:\\

``Sometimes physicists say that the electron's angular momentum and magnetic moment cannot be generated by rotation because the electron is too small: if the electron's radius is much smaller than the Compton radius, $\frac{\hbar}{mc}$, there is no way to generate an angular momentum of $\frac{\hbar}{2}$ without the electron's mass rotating faster than the speed of light and no way to generate a magnetic moment of $\frac{e \hbar}{2 m c}$ without the electron's charge rotating faster than the speed of light.  In brief, I take the solution to this puzzle to be that (in ordinary circumstances) the superposition of classical Dirac field states that forms the quantum state of the field is a superposition of states where the electron's relativistic mass (energy over $c^2$) and charge are not so tightly confined.  For example, in the hydrogen atom these might be states where the electron's relativistic mass and charge are spread throughout the atom's electron cloud---states where the electron is as big as the atom ({[Sebens, Charles T. 2021. \href{https://link.springer.com/article/10.1007/s10701-021-00480-7}{Electron charge density: A clue from quantum chemistry for quantum foundations}. {\em Foundations of Physics}, {\bf 51}, 75.]}, Sect.\ 4.4). It is possible to confine the electron's relativistic mass and charge so that they reside primarily within a sphere much smaller than the Compton radius, but it seems that when this is done the electron's relativistic mass becomes large and its magnetic moment becomes small, so that there is no need for either mass or charge to rotate superluminally (because there is enough relativistic mass to generate the ordinary angular momentum through rotation despite the small size of the mass distribution and because the rotation of charge does not have to yield the ordinary magnetic moment; {[Sebens, Charles T. 2020. \href{https://journals.aps.org/pra/abstract/10.1103/PhysRevA.102.052225}{Possibility of small electron states}. {\em Physical Review A}, {\bf 102}(5), 052225.]}).  That being said, if we define the velocity of energy (or relativistic mass) flow as the energy flux density ($c^2$ times the momentum density) over the energy density, then it will exceed the speed of light in certain circumstances ({[Bialynicki-Birula, Iwo \& Bialynicka-Birula, Zofia. 2022. \href{https://journals.aps.org/pra/abstract/10.1103/PhysRevA.105.036201}{Comment on ``Possibility of small electron states''}. {\em Physical Review A}, {\bf 105}(3), 036201.]}).  More research is needed to better understand the flow of energy in such situations and whether an always slower-than-light velocity of energy flow can be found.  For charge flow, this problem does not arise.  If we define the velocity of charge flow as the current density divided by the charge density, it cannot exceed the speed of light for any state of the classical Dirac field.''
\begin{flushright}
(This quote is from section 4.2 of: Sebens, Charles T. 2022.\\\href{https://link.springer.com/article/10.1007/s11229-022-03844-2}{The fundamentality of fields}. {\em Synthese}, {\bf 200}(5), 380.)
\end{flushright}

\vspace*{12 pt}Let me also take this opportunity to mention two other follow-up articles.  One article analyzes Stern-Gerlach electron spin measurements within classical Dirac field theory, depicting the evolution of a spinning electron:
\begin{quote}
Sebens, Charles T. 2021. \href{https://link.springer.com/article/10.1007/s11229-020-02843-5}{Particles, fields, and the measurement of electron spin}. {\em Synthese}, {\bf 198}(12), 11943--11975.
\end{quote}
Another article examines electric self-repulsion within the electron's charge distribution:
\begin{quote}
Sebens, Charles T. 2023. \href{https://link.springer.com/article/10.1007/s10701-023-00702-0}{Eliminating electron self-repulsion}. {\em Foundations of Physics}, {\bf 53}, 65.
\end{quote}

\end{document}